\documentclass[prd,showpacs,showkeys,preprint,nofootinbib]{revtex4}
\usepackage{color}
\usepackage{amsmath}
\usepackage{amssymb}
\usepackage{yfonts}[1988/10/03]
\usepackage{graphicx}
\usepackage{subfig}
\newcommand {\beq}{\begin{equation}}
\newcommand {\eeq}{\end{equation}}
\newcommand {\beqa}{\begin{eqnarray}}
\newcommand {\eeqa}{\end{eqnarray}}

\begin{document}\textsf{\today}  
\title{ Constraints on Dark Energy from the Gravitational Wave Background
}

\author{Jafar Khodagholizadeh }
\affiliation{Department of Physics Education,  Farhangian University, P.O. Box 14665-889, Tehran, Iran.
}
\begin{abstract}
Current observational data indicate that dark energy (DE) is a cosmological constant without considering its conclusiveness evidence. Considering the dynamic nature of $\Lambda$ individually as a function of time and the scale factor, we review their effects on the gravitational waves. This article is a continuation of the previous work \textit{JHEAp 36 (2022) 48-54}, in which DE only was based on Hubble's parameter and/or its derivatives. For the DE model based on the scale factor ($a^{-m}$), the results showed that the parameter $m$ is more limited as $ 2 < m \leqslant 3$ compared with the other models and due to the small value of DE density at the early universe. It is only in the mode $m=3$ that DE affects the low-frequency gravitational waves when its frequency is less than the $10^{-3}$Hz in a matter-dominated epoch. The broad bound on reducing the amplitude and the $ \textquoteleft\textquoteleft$B-B$\textquotedblright$ polarization multipole coefficients, from maximum to minimum, is for the models developed based on the Hubble parameter function.
There are primary sources of low- and very low-frequency GWs, such as the coalescence of massive black hole binaries with $M_{bh} > 10^{3} M_{sun}$, to determine the type of DE by mHz frequency space experiments (e.g., LISA) and by nHz-range NANOGrav 15-year data.
\keywords{ Dark energy models, NANOGrav 15-year data,  LISA }
\end{abstract}
\pacs{98.80.-k, 04.20.Cv, 02.40.-k}
%\preprint{***}
\maketitle
\section{Introduction}
Supernovae are strong evidence of accelerating the Universe expansion \cite{Riess,Perl,Amanullah}. Dark energy (DE) is considered the cause of this space expansion. DE measurements indicate that it contributes $68\%$ of the total energy in the present-day observable universe in the standard scenario of $\Lambda $CDM cosmology; however, its nature is very mysterious \cite{Lonappan:2017lzt}. There are various theoretical approaches to explain the role of DE  in general relativity, such as canonical scalar field, the so-called quintessence \cite{Fujii:1982ms}, a non-canonical scalar field (e.g., phantom \cite{Caldwell:1999ew}), tachyon scalar field motivated by string theories \cite{Padmanabhan:2009vy}, a fluid with a special equation of state (EoS) called as Chaplygin gas \cite{Kamenshchik:2001cp, Bento:2002ps, Bilic:2001cg}, and the holographic DE \cite{Li:2006ci,Nojiri:2006be}. The stellar remnant black holes and cosmologically coupled mass growth of them at redshifts below $2.5$ provide strong evidence for the astrophysical origin of DE \cite{Farrah:2023opk}. In addition, much of current observational data show that DE is a cosmological constant although there is no evidence for its conclusiveness\cite{126,128,130,132,133,134,136,137,143,145,146}.\\
A cosmological constant is mathematically, or perhaps physically, the simplest form of DE \cite{Einstein}, although its theoretical origin has not been understood yet\cite{Weinberg:1988cp}. The class of $\Lambda$ as a DE could be based on the Hubble parameter. The general form of this parameter is known as the generalized running vacuum model (GRVM): $\Lambda(H)=A+B H^{2}+ C\dot{H}$. Here, a dot denotes the differentiation with respect to cosmic time, $B$ and $C$ are constants and dimensionless, and $A$ has the unit of length$^{-2}$\cite{74}. This model has two sub-cases: the classical running vacuum model (RVM), obtained by setting C to zero \cite{24}, and the generalized running vacuum subcase (GRVS), where B is equal to 0 \cite{30}.
 In a recent study, the parameters B and/or C are constrained using data from SNeIa, cosmic chronometers, the CMB, and BAOs \cite{Farrugia:2018mex}. The next $\Lambda-$dynamics could be achieved as an explicit function of time and the form of the scale factor. The most popular explicit function of time has the form of the inverse power law as $\Lambda(t)\propto t^{-n}$ \cite{44,45,46,47,48}. Another function is proposed as an exponential decay \cite{52}. The last $\Lambda-$ model could be expressed in terms of the scale factor $a$, in which the general form is $ \Lambda(a)=A a^{n}+B a^{m}$ or the relation consistent with the data is $ \Lambda(a)=A + B a^{-m}$. Here, $m$ is a constant \cite{63}.\\
The North American Observatory for Gravitational Waves (NANOGrav) has recently provided evidence for a stochastic signal consistent with a stochastic gravitational waves background \cite{NANOGrav:2023gor}. Gravitational-wave observatories can be used to explore a wide range of fundamental physics problems throughout the universe's history, such as DE. Using the DE function in tensor mode perturbation, the values of $B$ and $C$ can be selected to effectively reduce the amplitude of primordial gravitational waves. Hence, all the quadratic effects of the tensor modes in the CMB (e.g., tensor contribution to the temperature multipole coefficients $C_{l}$) and all of the $ \textquoteleft\textquoteleft$B-B$\textquotedblright$ polarization multipole coefficients are maximum $60\%$ less than that in the case without the damping due to DE terms with total vacuum contribution \cite{Khodagholizadeh:2022ldk}. In addition, the maximum reduction for polarization multipole coefficients is observed in the GRVM, which is $42\%$.\\
The field theory of DE affects gravitational wave propagation, depending on time, frequency, and polarization \cite{Romano:2022jeh}. This effect is contrary to the previous cases where DE itself is the source of gravitational waves. Therefore, interactions of DE fields with the rest of the universe extract the gravitational wave signal with an amplitude $ \sim 10^{-2} $, which corresponds to a frequency of $ \sim 10^{-13} Hz $ \cite{Jhalani:2017rgf,Anupam}. The present article aims to study the effect of DE on gravitational waves based on the other mentioned models and compare the obtained results. To this end, we assumed that the equation-of-state parameter of DE, $w$, is fixed at -1, similar to $\Lambda$CDM.\\
The remainder of this paper is organized as follows: In Sec.II, we review briefly the equation of the tensor mode perturbation in the presence of $\Lambda$ as dynamical DE. In Sections II and III, the solutions of the field equations are studied for two DE models as functions of time and the scale factor, respectively. Finally, the results will be reported.

\section{ Linear field equation with dynamic dark energy} 
In the following, the perturbed metric is decomposed as:
\begin{eqnarray}
	g_{\mu\nu}=\bar{g}_{\mu\nu}+h_{\mu\nu},
\end{eqnarray}
where $\bar{g}_{\mu\nu}$ is the background metric and $h_{\mu\nu}$ is the symmetric perturbation term with the condition $\mid h_{\mu\nu}\mid\ll 1 $. The metric components of the Friedmann-Lemaitre-Robertson-Walker (FLRW) model in the Cartesian coordinate system are \cite{1}:
\begin{eqnarray}\label{metric}
	\bar{g}_{00}&=&-1,\;\;\;\;\bar{g}_{i0}=0,\;\;\;\bar{g}_{ij}=a^2(t)\tilde{g}_{ij}
	,\;\;\;\tilde{g}_{ij}=\delta_{ij}+K\frac{x^ix^j}{1-Kx^2},
\end{eqnarray}
where $i$ and $j $ rims over the values 1, 2, and 3; $x^0=t $ is the time coordinate in our units, $K$ is the curvature constant, and the speed of light is equal to 1. Also, $a(t)$ is the scale factor, which will be $\alpha\cosh(t/\alpha)$ and $ \alpha=\sqrt{\dfrac{3}{\Lambda}}$ in the closed de Sitter spacetime, which is the maximally extended spacetime. From \cite{Pad}, the field equation for the tensor mode fluctuation in the source-free region is:
\begin{eqnarray}\label{Eq1}
	\square h_{\mu\nu}+2 \bar{R}_{\mu\alpha \nu\beta}^{0} h^{\alpha\beta}=0 
\end{eqnarray}
where $\bar{R}_{\mu\alpha \nu\beta}^{0} $ is the background Riemann tensor. This equation describes the propagation of weak GWs in the source-free region of the curved spacetime. 
By using the Friedmann equation, $ 2\dot{a}^2+a\ddot{a}=\Lambda(t) a^2-2K$, while the nature of cosmological constant is dynamic. Also, from $ h_{ij}=a^{2} D_{ij} $, Eq. (\ref{Eq1}) will be \cite{Khodagholizadeh, Khodagholizadeh:2014ixa}: 
\begin{eqnarray}
	\nabla^2D_{jk}-a^{2}\ddot{D}_{ij}-3a\dot{a}\dot{D}_{ij}+(\Lambda(t) a^{2}-a\ddot{a}-2K) D_{ij}=0
\end{eqnarray}
Using the method of separation of variables $D(z,t)=\hat{D}(z) D(t)$, the time evolution of the tensor mode perturbation will be as follows (for more details, see \cite{Khodagholizadeh:2022ldk}):
\begin{eqnarray}\label{final}
	\ddot{D}(t)+ 3 \dfrac{\dot{a}}{a}\dot{D}(t)+\dfrac{q^{2}}{a^{2}}D(t)=(\Lambda(t) - \dfrac{\ddot{a}}{a})D(t)
\end{eqnarray}
The expression $q^{2}= n^{2}-3+2K $ is also a wavenumber. Both the nonzero spatial curvature parameter of the background and the cosmological constant are in the field equation of tensor perturbation. Meanwhile, the low-frequency gravitational waves with nonzero background curvature alone produce interesting results so that $n$ will be a discrete number \cite{Khodagholizadeh:2021kfy}. It is noticed that the nonzero curvature affects constraining some DE models \cite{Polarski, Franca, Ichikawa, Ichikawa1, Clarkson, Gong, Ichikawa2, Wright, Zhao}. Some data such as CMB, type Supernova Ia (SNe Ia), and galaxy survey show that if DE density is free of redshift, the bounds on cosmic curvature are less stringent depending on the assumption about its early time properties. However, assuming a constant DE equation of state gives the most stringent constraints on cosmic curvature\cite{Wang:2007mza}. 
It seems that the presence of curvature will eventually cause a phase difference in the wave number, and this difference should be considered only in low continuous wave numbers. The curvature of the background spacetime in the presence of neutrinos and anti-neutrinos is effective on the amplitude reduction of the primordial gravitational waves \cite{Khodagholizadeh:2014ixa,Khodagholizadeh:2017ttk, khoda}.  
However, regardless of whether the wave number is discrete or continuous, the time evolution of gravitational waves depends on the value of $n$, so from now on we can introduce $D(t)$  as $D_n(t)$.

\section{ Dark energy as a function of Time}
The dynamic nature of $ \Lambda $ could be conventionally an explicit function of time, in which the most popular relation is in the form of the inverse power law as $ \Lambda(t)=\Lambda_{bare}+ \dfrac{\alpha^{2}}{ t^{2}} $. Here, $ \Lambda_{bare} $ is a constant limit of $ \Lambda(t) $ as a $ a\longrightarrow \infty $ and $ \alpha^{2} $ is a real parameter (either positive or negative )\cite{Szydlowski:2015rga}. This model interprets running $ \Lambda(t) $ as a special model of interacting cosmology with the interaction term, $ -d\Lambda(t)/dt $, where energy transfer is between dark matter and DE sectors.  Since we are not concerned with inflation epoch and even earlier but rather with the late time behavior of dark energy models, so we can neglect this parameter \cite{46, Lima:2015kda,Urbanowski:2012pka,Wang:2004cp}. The best limit on the value of parameter $ \alpha^{2} $ using the astronomical data such as SNIa data, BAO, and the CMB is $ -0.014< \alpha^{2}< 0.012 $ (for $95\%$ CL)\cite{Szydlowski:2015rga}. In the next step, by choosing the largest value of $\alpha^{2}$, regardless of whether it is negative or positive, we will examine the effect of this case on GWs in the early universe. 
\subsection{ Damping effect in the Early Universe} 
By using the Friedmann equation $\dfrac{8\pi G \bar{\rho}}{3}=H^{2}=\dfrac{1}{4t^{2}}$, and $a(t)= t^{1/2}$ as the scale factor in the radiation-dominated era, Eq. (\ref{final}) becomes
\begin{eqnarray}\label{final1}
	\ddot{D}_n(t)+ 3 \dfrac{\dot{a}}{a}\dot{D}_n(t)+\dfrac{q^{2}}{a^{2}}D_n(t)=\dfrac{4 \alpha^{2} +1 }{4 t^{2}}D_n(t)
\end{eqnarray}
The treat of the tensor mode perturbation in the radiation- and matter-dominated eras can be investigated by replacing the independent variable $t$ with $u= q\tau= q\int_{0}^{t} \dfrac{dt^{\prime}}{a(t^{\prime})}=\dfrac{2qt}{a(t)}$. For this purpose, we consider the largest value of $ \alpha^{2} $, i.e., $ -0.014 $. Therefore, this equation will be 
\begin{eqnarray}\label{224}
	\dfrac{d^{2}}{du^{2}} D_{n}(u)+ (\dfrac{2}{u}) \dfrac{d}{du} D_{n}(u)+D_{n}(u)=(\dfrac{0.944 }{u^{2}}) D_{n}(u)
\end{eqnarray}
Generally, tensor mode perturbation rapidly became time-independent after horizon exit and remains as such until horizon re-entry; thus, initial conditions are as follows:
\begin{eqnarray}
	D_{n}(0)=1~~~~~~,~~~~~~\dfrac{d}{du}D_{n}(0)=0
\end{eqnarray}
For deep inside the horizon, Eq. (\ref{224}) approaches a solution as:
\begin{eqnarray}
	D_{n}(u)=D_{n}^{0} j_{_{(0.5927)}}(u) + D_{n}^{1} y_{_{(0.5927)}}(u)
\end{eqnarray}
where $j_{_{(0.5927)}}(x) $ and $ y_{_{(0.5927)}}(x) $ are the spherical Bessel functions and $ D_{n}^{0} $ and $ D_{n}^{1} $ are constant parameters. Using their definition, the general solution is:
\begin{eqnarray}
	D_{n}(u)&=&[D_{n}^0 + \dfrac{0.124}{u^{2}}D_{n}^0 +\dfrac{0.47}{u}D_{n}^{1}] \dfrac{\cos(2.501-u)}{u}-[D_{n}^1 - \dfrac{0.472}{u}D_{n}^0 +\dfrac{0.124}{u^{2}}D_{n}^{1}] \dfrac{\sin(2.501-u)}{u}\nonumber\\
\end{eqnarray}
For large $u (u\gg 1)$, the tensor modes are deep inside the horizon, and the solution approaches the homogeneous solution. Therefore, the coefficient of the $\dfrac{\cos u}{u} $ term must be 0 and the coefficient of the $\dfrac{\sin u}{u} $ term must be equal to 1. As a result, the constant values will be $ D_{n}^0= \sin(2.501)$ and $ D_{n}^1= cos(2.501)$. In addition, compared with the solution $\dfrac{\sin u}{u}$, in the absence of DE, Eq. (\ref{224}) shows that $ D_{n}(u)$ follows the without DE solution rather accurately until $ u\approx 1$. When the perturbation enters the horizon and, thereafter, it rapidly approaches $ \approx \dfrac{\sin(u+\delta)}{u}$, in which $ \delta$ is very small and negligible. Therefore, the numerical solution shows that time-dependent DE does not reduce the squared amplitude for wavelengths that enter the horizon during the radiation-dominated phase, independent of any cosmological parameters. This result was also obtained in the case of the cosmological constant based on a function of Hubble's parameter and/or its derivatives \cite{Khodagholizadeh:2022ldk}. However,  other reasons should also be sought for this issue.
\subsection{ Short wavelengths in a matter-dominated era}
Using the definition $u=q\tau=q\int_{0}^{t}\dfrac{dt^{'}}{a(t^{'})}=\dfrac{3qt}{a(t)}$, where $a(t)= t^{2/3}$ is the scale factor in the matter-dominant era, the equation of the tensor mode evolution will be: 
\begin{eqnarray}\label{115}
	\dfrac{d^{2}}{du^{2}} D_{n}(u)+ (\dfrac{4}{u}) \dfrac{d}{du} D_{n}(u)+D_{n}(u)=(\dfrac{1.874}{u^{2}}) D_{n}(u)
\end{eqnarray}
The general solution is based on the Bessel functions, as:    
\begin{eqnarray}
	D_n(u)=  \dfrac{1}{u^{3/2}}[D_{n}^{0}J_{2.03}(u)+ D_{n}^1 Y_{2.03}(u)]
\end{eqnarray}
where $D_{n}^{0}$ and $D_{n}^{0}$ are constant. Thus, for large $ u $, the above solution approaches to 
\begin{eqnarray}
D_{n}(u)&=&\lbrace [0.797 D_{n}^0 +1.545\dfrac{D_{n}^1}{u} -0.724 \dfrac{D_{n}^0}{u^{2}}] \cos(3.975)\nonumber\\ &+&[-0.797 D_{n}^1 +1.545\dfrac{D_{n}^0}{u} +0.724 \dfrac{D_{n}^1}{u^{2}}] \sin(3.975)\rbrace\dfrac{\cos u}{u^{2}} \nonumber\\ &+&\lbrace [0.797 D_{n}^0 +1.545\dfrac{D_{n}^1}{u} -0.724 \dfrac{D_{n}^0}{u^{2}}] \sin(3.975)\nonumber\\ &-&[-0.797 D_{n}^1 +1.545\dfrac{D_{n}^0}{u} +0.724 \dfrac{D_{n}^1}{u^{2}}] \cos(3.975)\rbrace\dfrac{\sin u}{u^{2}}
\end{eqnarray}
 Deep inside the horizon, when $u\gg 1$, the right-hand side of Eq. (\ref{115}) becomes negligible, and the solution approaches a homogeneous solution as $\dfrac{\sin u}{u^2}$. Therefore, the coefficient of the first term, $ \dfrac{\cos u}{u^{2}} $ must be 0 and the coefficient of the second term, $ \dfrac{\sin u}{u^{2}} $, must be 1. Compared with the solution $\dfrac{\sin u}{u^2}$ in the absence of DE, Eq. (\ref{115}) shows that $ D_{n}(u)$ follows the DE-free solution rather accurately until $ u\approx 1$. When the perturbation enters the horizon and, thereafter, rapidly approaches $ \approx 0.797 \dfrac{\sin(u+\delta)}{u^2}$, in which $ \delta$ is very small and negligible (see Fig. \ref{L0}, solid line (gray)). Furthermore, it has a significant effect on decreasing the $ \textquoteleft\textquoteleft$B-B$\textquotedblright$ polarization multipole coefficient, $ C_{lB}$, which is up to $36.4\%$ less than that without the damping due to DE, which is a function of time.\\
 The noteworthy point is that when we consider the upper limit, $ \alpha^{2}< 0.012 $, it will not be much different. The coefficient $ \alpha^{2} $ will only cause some disturbance to the spherical Bessel functions as $ j_{n+\delta} (x) $ as long as the $ \delta $ is less than $ 0.01 $, e.g., $ \delta< 0.1 $. Hence, the results will not be changed, and the same effect happens during the matter-dominated era.
\subsection{ General wavelengths in the dark energy-dominated era} 
The universe begins to accelerate after the redshift is nearly less than $0.5$ while the energy density of vacuum, $ \Omega_{\Lambda} $, is in the range of $ 0.5 $ to $ 1 $. This accelerated expansion of the universe was discovered for the first time by using distant of type Ia supernovae \cite{Frieman:2008sn, Broadhurst:1988ym,Peebles:2002gy}. The evolution of gravitational wave at present epoch can be assessed by replacing the independent variable $ t$ with $ \chi =\dfrac{\bar{\rho}_{\Lambda}}{\bar{\rho}_{M}}= \dfrac{\bar{\rho}_{\Lambda, EQ}}{\bar{\rho}_{M,EQ}}\dfrac{a^{3}}{a_{EQ}^{3}}$. Here, $a_{EQ}$, $\bar{\rho}_{\Lambda, EQ}$ and $\bar{\rho}_{M, EQ} $ are the value of the Robertson-Walker scale factor, energy densities of vacuum and matter at matter-vacuum equality. According to the Friedmann equation, we have:
\begin{eqnarray}
	H_{EQ} \dfrac{dt}{\sqrt{2}}=\dfrac{d\chi}{3\chi\sqrt{(1-\Omega_{K,EQ})(\chi+\chi^{2})+2 \Omega_{K,EQ}\chi^{4/3} }}
\end{eqnarray} 
where $\Omega_{K, EQ}$ is the curvature density at matter-vacuum equality. When the DE is important, e.g., $\chi \gg 1 $, the homogeneous equation of GWs in the expansion universe will be as follows:
\begin{eqnarray}\label{30}
	\dfrac{d^{2}}{d\chi^{2}}D_{n}(\chi)+ \dfrac{2}{\chi} \dfrac{d}{d\chi} D_{n}(\chi)+\dfrac{\kappa^{'^{2}}}{\chi^{8/3}}D_{n}(\chi)=0
\end{eqnarray} 
where $ \kappa^{'}$ is the dimensionless rescaled wavenumber as $\kappa^{'^{2}} =\dfrac{2q^{2}}{9 H_{EQ}^{2} a_{EQ}^{2}(1-\Omega_{K,EQ})}$. The propagation of GWs at vacuum epoch depends on the density of curvature energy as $\Omega_{K,EQ}=(1+0.5)^{2}\Omega_{K}^{0}$ with $\Omega_{K}^{0}=-0.0001_{-0.0052}^{+0.0054}$\cite{Planck:2015fie}. However, it can be ignored due to its very small value. Regardless of the value of $ \kappa^{'}$, the general solution of Eq.(\ref{30}) will be  
	\begin{equation}
D_{n}(\chi)= D_{n}^{0} \dfrac{3 \kappa^{'}}{\chi^{1/3}} \sin(\dfrac{3 \kappa^{'}}{\chi^{1/3}})  + D_{n}^{1}\cos(\dfrac{3 \kappa^{'}}{\chi^{1/3}})
	\end{equation}
	Here, $ D_{n}^{0} $ and $ D_{n}^{1} $ are constant. There will be no general solution when $ \kappa^{'} \ll 1$ and $ \chi\gg 1$ separately and under the condition $ \dfrac{\kappa^{'}}{\chi^{4/3}}=\dfrac{q}{a} \longrightarrow 0$ the solution tends to $ D_{n}^{1} $. In the cases of long wavelengths, $\kappa^{'}\ll 1$ and $\kappa^{'}\gg 1$ correspond to the short wavelength. The cosmological gravitational waves are detectable when $\kappa^{'}\gg 1$, which leads to $ D_{n}^{1}=0 $.
	Therefore, since the damping effect is small in any way for $ \kappa^{'}\ll 1$, it will be an adequate approximation for all the wavelengths to take the solution of Eq. (\ref{30}) in the $\Lambda$-dominated era by multiplying a factor $ \xi(\kappa^{'}) =\dfrac{1+ 0.797\kappa^{'}}{1+\kappa^{'}}$ as 
\begin{eqnarray} \label{177}
	D_{n}(\chi)=\xi(\kappa^{'}) \dfrac{3 \kappa^{'}}{\chi^{1/3}} \sin(\dfrac{3 \kappa^{'}}{\chi^{1/3}})
\end{eqnarray} 
Thus, $ \xi(\kappa^{'}) $ is the reduction amplitude factor. For, $ \kappa^{'}\ll 1$, we have $\xi(\kappa^{'})=1$ and when $\kappa^{'}\gg 1$, the above relation will be:
\begin{eqnarray}
	D_{n}(\chi)=0.797\dfrac{3 \kappa^{'}}{\chi^{1/3}} \sin(\dfrac{3 \kappa^{'}}{\chi^{1/3}})
\end{eqnarray}
Here, again we introduce a dimensionless rescaled wave number 
\begin{eqnarray}\label{ka}
	\kappa^{'}=\dfrac{q\sqrt{2}}{a_{EQ}H_{EQ}}=\dfrac{(q/a_{0})}{3H_{0}\Omega_{M}}(\dfrac{\Omega_{M}^{4}}{\Omega_{\Lambda}})^{1/6}=\dfrac{2831(q/a_{0})}{\Omega_{M}h^{2}}[Mpc^{-1}]
\end{eqnarray} 
\begin{figure} 
	{	\includegraphics[width=0.49\textwidth]{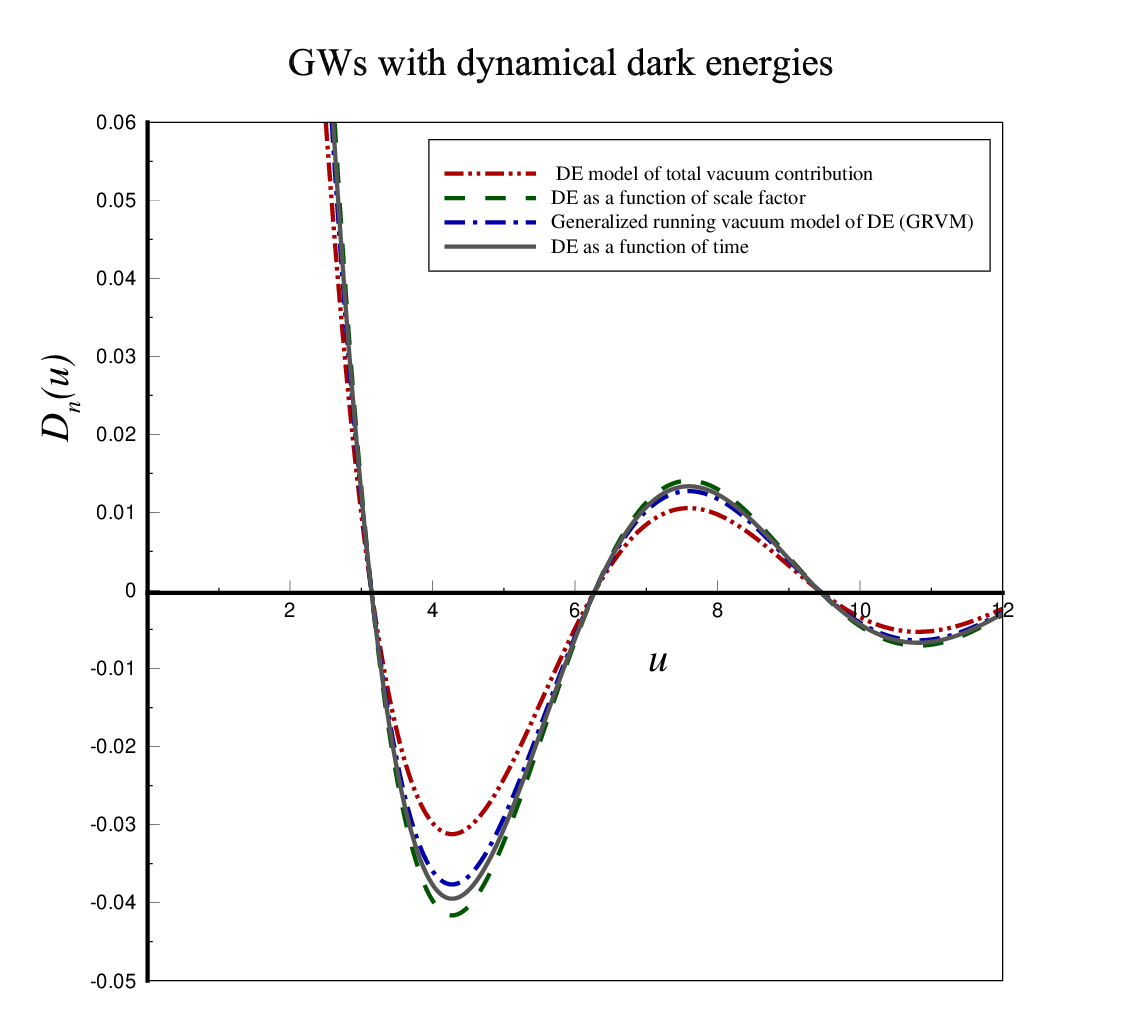} 
	}\hfill 
	{ 	\includegraphics[width=0.49\textwidth]{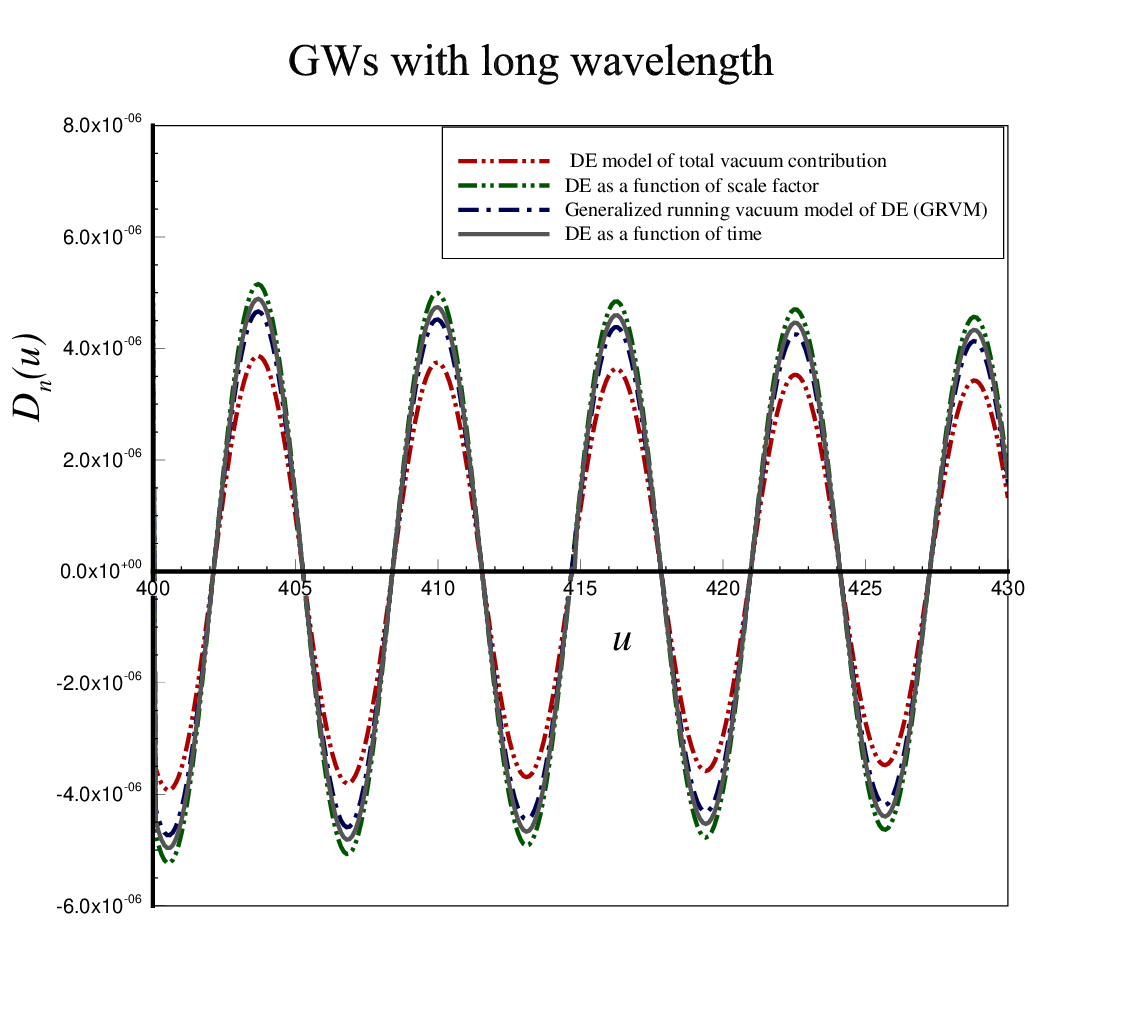} 
	} 
	\caption{Comparing GWs background in the presence of different dynamic dark energies in the matter-dominated era shows that at the beginning of entering the horizon (left plot), the differences are more obvious than they are deep inside of the horizon (right plot) in the long time propagation of gravitational waves background. This phenomenon can be studied by the NANOGrav 15-year data set. The solid (gray) and dashed (green) lines are for GWs in the presence of DE as a function of time and scale factor, respectively. GWs with DE based on the Hubble parameter are denoted with a dashed-dot-dot (red) line for total vacuum contribution and a dashed-dot (blue) line for the GRVM model.}
	\label{L0}
\end{figure}
where $a_{0}$ is the present-day scale factor in which we used the relations $ H_{EQ}=\sqrt{2}(H_{0}\sqrt{\Omega_{M}}) (a_{0}/a_{EQ})^{3/2} $ and $ (a_{0}/a_{EQ})^{3}=\Omega_{\Lambda}/\Omega_{M} $. Also, $H_{EQ}=H_{0}\sqrt{2\Omega_{M}}(1+z_{EQ})^{3/2}$ and $z_{EQ}$ are the Hubble rate and the redshift at matter-vacuum equality, respectively. In addition, we have $ \dfrac{\Omega_{\Lambda}}{\Omega_{M}}=(1+z_{EQ})^{3}$. As mentioned, the important point in detecting the cosmological gravitational waves is $\kappa^{'}\gg 1$, which is independent of any cosmological parameters such as $a_0$ and $\Omega_{M}$. %. Overall, it is $10^3$ order larger, and it is big enough.\\
According to the relation (\ref{ka}), a gravitational wave with a frequency of $qc/2\pi a_{0}=10^{-4}$ Hz would have $\kappa^{'}= 5.6 \times 10^{15}/\Omega_{M}h^{2} \gg 1$ \cite{Komatso}. 
For smaller frequencies, e.g., $10^{-7}$Hz, the value of $\kappa^{'}$ will be $\kappa^{'}= 5.6 \times 10^{12}/\Omega_{M}h^{2} \gg 1 $, which is still a very large number that guarantees the presence of DE along with gravitational waves. It is noticed that the maximum acceptable frequency to observe this effect is $10^{-4}$Hz. \\
The ground-based interferometers are sensitive to around $100$ Hz. Also, it is reported that GWs are coming from compact binaries, supernovae, and pulsars \cite{GWs}. Thus, it is not sensitive to detect cosmological GWs. Moreover, due to unavoidable seismic noise, low-frequency GWs in the frequency range lower than 1 Hz are not detectable on Earth. Hence, this model of DE effect could be detected by the space borne future interferometer of GWs such as pulsar timing arrays (PTA), which operate at frequencies around $10^{-7}$ to $10^{-8}$ Hz. The PTAs use shifts in the beating of observed pulsar beams as gravitational waves stretch and squeeze the space between the Earth and pulsars. As mentioned, NANOGrav \cite{NANOGrav:2023gor,NANOGrav:2023hvm}, The European Pulsar Timing Array (EPTA \cite{Antoniadis:2023lym,Antoniadis:2023ott}), the Parkes Pulsar Timing Array in Australia (PPTA \cite{Zic:2023gta,Reardon:2023gzh}), and the Chinese Pulsar Timing Array (CPTA \cite{Xu:2023wog}) have provided evidence for a stochastic signal consistent with a SGWB to prove the imprints of SMBHBs. This signal had existed at the early stages of the universe with the mass of around $10^{2}-10^{11} M_{sun}$ at the redshift $ 10< z< 15$ \cite{NANOGrav:2023pdq,Banados:2017unc}. Detection of stochastic gravitational waves background (SGWB) brings us a great opportunity to prove the imprints of DE in the long wavelengths or low-frequency in the DE-dominated (see the right plot in Fig. \ref{L0}). 
\section{ Dark energy as a function of scale factor }
From the principle of quantum cosmology, it is convenient to use the scale factor instead of time for the cosmological constant with no conflict with observation\cite{Ozer:1985ws}. This theory was first proposed by M. Gaspirini in a thermal approach \cite{Gasperini:1987fi}. Such dependence also appears in string-dominated cosmology\cite{Vilenkin} and semiclassical Lorentzian analysis of quantum tunneling \cite{Strominger:1988yt}. Here, a class of $ \Lambda $ decreases as $ a^{-m} $, where $ a $ is the scale factor of the Friedman-Robertson-Walker (FRW) metric and $ m $ is a constant parameter in which $ 0\leq m \leq 3$\cite{Silveira:1994yq}. The constraints imposed on the models by the magnitude-redshift test show a model with $ m\gtrsim 1.6 $ in good agreement with the data. Here, high-redshift type Ia supernovae (SNe Ia) are used as standard candles \cite{Silveira:1997fp}. Although Chen and Wu \cite{Chen:1990jw} gave some interesting arguments favoring the special value $ m = 2$, it is clear that the functional dependence with the scale factor is only phenomenological and does not come from particle physics first principles. Indeed, this effect can be seen on a large scale. For instance, the angular power spectrum (up to $\ell= 20$) for different values of $m$ and $ \Omega_{m,0} $ has been computed by studying the effect of the $ \Lambda $- decaying term on the cosmic microwave background anisotropy. \\
With this form of DE in the early universe epoch, we consider first wavelengths short enough to have re-entered the horizon during the radiation-dominated era. Thus, using the Friedmann equation with the scale factor $a(t)= \sqrt{t}$, and the independent variable $u= q\tau= q\int_{0}^{t} \dfrac{dt^{\prime}}{a(t^{\prime})}=\dfrac{2qt}{a(t)}$ instead of $t$, Eq. (\ref{final}) becomes:
\begin{eqnarray}\label{}
	\dfrac{d^{2}}{du^{2}} D_{n}(u)+ (\dfrac{2}{u}) \dfrac{d}{du} D_{n}(u)+D_{n}(u)=(2^{m-2}\dfrac{ q^{m-4}}{u^{m-2}}+\dfrac{1}{u^{2}}) D_{n}(u)
\end{eqnarray}
 \begin{figure} 
	{	\includegraphics[width=0.48\textwidth]{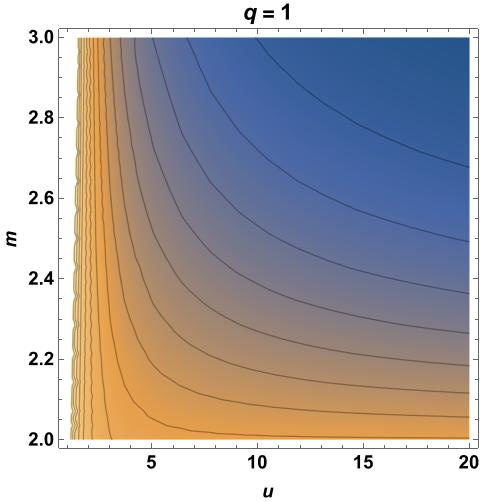} 
	}\hfill 
	{ 	\includegraphics[width=0.48\textwidth]{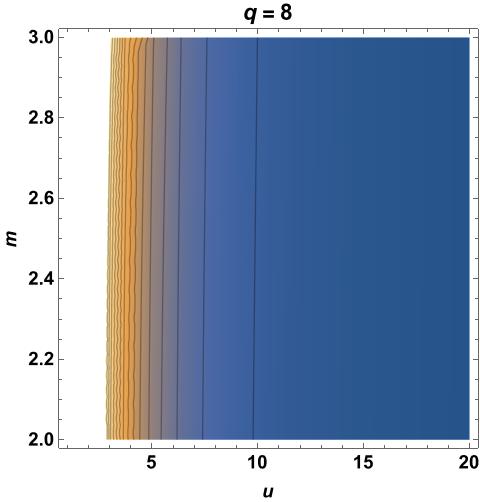} 
	} 
	\caption{ Density-plot are drawn based on specific wave numbers. Left: by choosing the wave number $q=1$, the interval of $m$ has become smaller and the probability of the presence of the wave in all u 's is very low. Right: in the next wave numbers, these divergence restrictions have been removed.~~~~~~~~~~~~~~~~~~~~~~~~~~~~~~~~~~~~~~~~~~~~~~~~~ } 
	\label{L1}
\end{figure}
\begin{figure} 
	{	\includegraphics[width=0.490\textwidth]{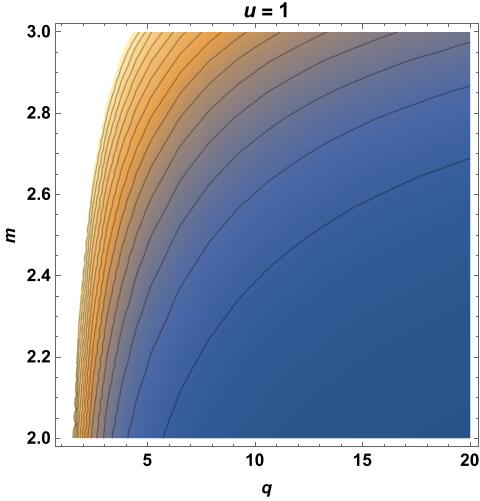} 
	}\hfill 
	{ 	\includegraphics[width=0.49\textwidth]{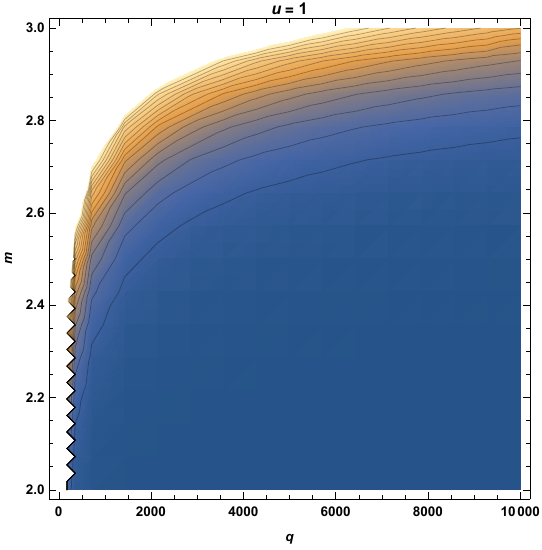} 
	} 
	{	\includegraphics[width=0.49\textwidth]{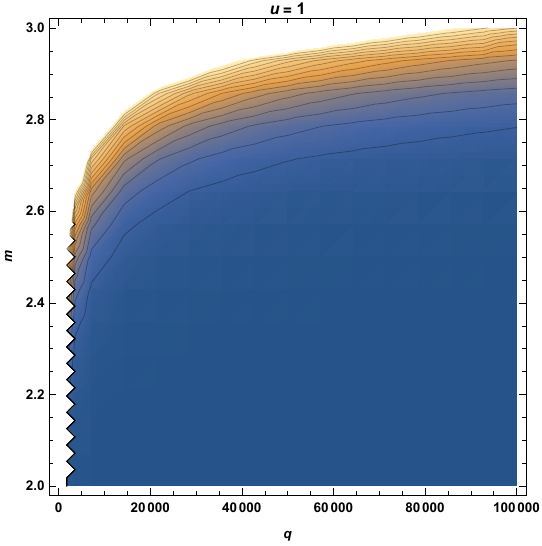} 
	}\hfill 
	{ 	\includegraphics[width=0.49\textwidth]{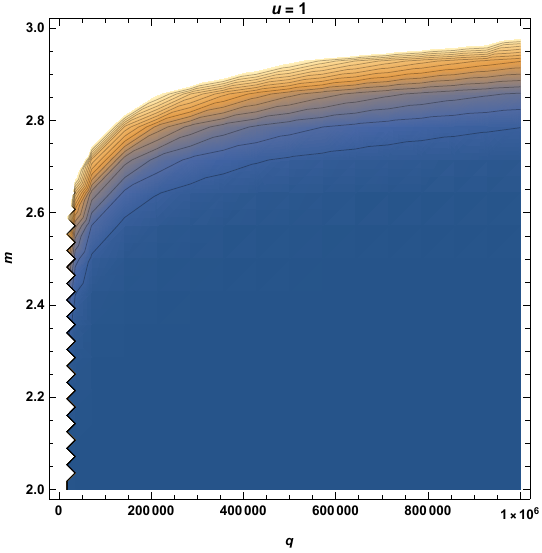} 
	} 
\caption{  Density plots are drawn for different wavenumbers in a specific interval of $m$ when the tensor mode perturbation enters the horizon in $u=1$. In low wave number (top-left), the interval is valid. However, as the wave number increases, the probability of the presence of the interval decreases (top-right and bottom-left). Finally, at the wave number of $q=10^{6}$ and above, the value of $m$ is fixed at 2.8 ( bottom-right) in $q\lesssim 10^{5}$, which corresponds to about $10^{-3}$Hz.~~~~~~~~~~~~~~~~~~~~~~~~~~~~~~~~~~~~~~~~~~~~~~~~~~~~~~~~~~~~~~~~~~~~~~~~~} 
	\label{L2}
\end{figure}
The influence of the DE in the early universe seems very small compared to its late-time value. The power measurement of small-scale CMB temperature anisotropy from the South Pole Telescope (SPT) bandpowers improves the early DE density such that $ \Omega_{e} $ is reduced from $0.052$ for WMAP7-only to $0.013$ for WMAP7+SPT. This is a $38\%$ improvement on the upper limit of $\Omega_{e} < 0.018$ reported for WMAP7+K11 \cite{Reichardt}. The upper limit is essentially unchanged at $ \Omega_{e} < 0.014 $ for WMAP7+SPT+BAO+Sne. However, the $\Omega_{e} < 0.013 $ bound from WMAP+SPT is still the best-published constraint from the Cosmic Microwave Background Radiation\cite{Hou2014}. Therefore, due to the weak existence of the DE in the early universe, it might not affect even the evolution of gravitational waves. This result was also obtained in the previous work according to the total vacuum contribution model, assuming the simplest form of DE coupled to the matter \cite{Khodagholizadeh:2022ldk}. Furthermore, for the previous model, DE was proportional to explicit function of time and affects gravitational waves during the radiation-dominated era. \\ Accordingly, for large $ u $ $ (u\gg 1)$, when the tensor modes are deep inside the horizon, the coefficient of the right-hand side of the above equation must be 0:
\begin{equation}
2^{m-2}\dfrac{ q^{m-4}}{u^{m-2}}+\dfrac{1}{u^{2}}=0
\end{equation}
Thus, for the general solution corresponding to the homogeneous equation state, $m$ must be greater than $2$, i.e., $ m> 2 $. In this specific interval, the wave number $q$ should be considered greater than $1$ (see Fig. \ref{L1} ). In $ u \approx 1 $, when the tensor modes enter the horizon, the right-hand side will be $ 2^{m-2} q^{m-4}+1=0 $, which is a coupled relation between wave number $ q $ and parameter $m$. This interval is somewhat valid for low frequencies, but it decreases for greater frequencies (see Fig. \ref{L2}). It can be mentioned that in a limited range of higher frequencies, the gravitational waves do not understand this case of DE for $a^m$ with power greater than $2.8$. As a result, this coefficient is more limited, i.e., $ 2< m \lesssim 2.8$ (see Fig. \ref{L2}, bottom-right).  
\subsection{ 	Short wavelength in the matter-dominated era   }
To deal with perturbations that may enter the horizon after the matter-energy density has become important, let us switch from $t$ to the independent variable $u=q\tau=q\int_{0}^{t}\dfrac{dt^{'}}{a(t^{'})}=\dfrac{3qt}{a(t)}$. Here, $a(t)= t^{2/3}$ is the scale factor in the matter-dominant era. Thus, the equation of the tensor mode time evolution with DE will be: 
\begin{eqnarray}\label{115}
	\dfrac{d^{2}}{du^{2}} D_{n}(u)+ (\dfrac{4}{u}) \dfrac{d}{du} D_{n}(u)+D_{n}(u)=(3^{2m-4}{\frac{q^{2m-6}}{u^{2m-4}}}+\dfrac{2}{u^{2}}) D_{n}(u)
\end{eqnarray}
This equation is very complicated because it depends on the two independent parameters wave number, $q$ and $ m$. Therefore, it cannot be solved easily. According to the previous results, the solution of the tensor mode perturbation field equation in the matter-dominated era was in terms of the Bessel functions. Thus, the suggested general form of the solution is as follows:
\begin{eqnarray}\label{general}
D_n(u)= A(u,q,m)
{{\rm J}_{n}\left({u}\right)}
+ B(u,q,m)
{{\rm Y}_{n}\left({u}\right)}
\end{eqnarray}
where ${{\rm J}_{n}\left({u}\right)}$ and $ {{\rm Y}_{n}\left({u}\right)}$ are Bessel functions. $A(u,q,m)$ and $B(u,q,m)$ are unknown functions. By inserting this solution in Eq. (\ref{115}) and setting unrelated coefficients equal to 0, equations for undefined functions $A(u,q,m)$ and $B(u,q,m)$ are obtained as follows:
\begin{eqnarray}
&&2 u \dfrac{dA(u,q,m)}{du}+3 A(u,q,m)=0 \nonumber\\
&& 2u \dfrac{dB(u,q,m)}{du}+3 B(u,q,m)=0
\end{eqnarray}
The solutions of this equation will be $A(u,q,m)=\dfrac{D_{n}^{0}}{u^{3/2}}$ and $ B(u,q,m)=\dfrac{D_{n}^{1}}{u^{3/2}}$, where $D_{n}^{0}$ and $D_{n}^{1}$ are constants. Again, putting these answers into the original equation gives:
\begin{eqnarray}
\dfrac{u^{\frac{1}{2}-m}}{q}\left[4\times9^{m}q^{2m}u^{6}-81 q^{6} u^{2m}(-17+4n^{2})\right] \left( D_{n}^{0}
{{\rm J}_{n}\left({u}\right)}
+ D_{n}^{1}
{{\rm Y}_{n}\left({u}\right)}\right)=0
\end{eqnarray}
With the coefficient becoming 0, it leads to $m=3$ and $n=\sqrt{\frac{53}{2}}$. Therefore, the only nonzero solution will be in low-frequency ($q \leq10^{5}$) (see Fig.\ref{L2}, bottom-left and -right ) as: 
\begin{eqnarray}\label{general}
D_n(u)= \dfrac{D_{n}^{0}}{u^{3/2}}
{{\rm J}_{\frac{\sqrt{53}}{2}}\left({u}\right)}
+ \dfrac{D_{n}^{1}}{u^{3/2}}
{{\rm Y}_{\frac{\sqrt{53}}{3}}\left({u}\right)}
\end{eqnarray}
For a large $u$, it takes the following form: 
\begin{equation}
 D_n (u)=\dfrac{D_n^{0}}{\sqrt{2\pi}}(2 \sin \gamma - 13\dfrac{\cos \gamma}{u})\dfrac{\sin u}{u^{2}}
\end{equation}
where $\gamma=\dfrac{(1+\sqrt{53})\pi}{4}$. A numerical method shows that $ D_n (u) $ follows the zero DE solution pretty accurately until $ u\approx1 $, when the perturbation enters the horizon. Afterward, it sharply approaches the $ 0.84 \dfrac{\sin u}{u^{2}} $ such that the damping effect of the nonzero DE reduces the tensor amplitude by the same factor $ 0.84 $. As a result, the tensor contribution to the temperature multiple coefficients $ C_{\ell} $ and the $ \textquoteleft\textquoteleft$B-B$\textquotedblright$ polarization multiple coefficients $ C_{\ell B} $ are $ 30\% $ less than they would be without damping due to this case of DE model with $ m=3$. This wave amplitude reduction is similar to models RVM and GRVS, which are $ 0.83 \dfrac{\sin u}{u^{2}} $ and $ 0.85 \dfrac{\sin u}{u^{2}} $, respectively \cite{Khodagholizadeh:2022ldk}.\\
In general, when the perturbation enters the horizon in expansion era, it is sufficient to reconsider Eq. (\ref{177}) with new reduction amplitude factor as $\xi(\kappa^{'}) =\dfrac{1+ 0.84\kappa^{'}}{1+\kappa^{'}}$ for the condition $\kappa^{'}\gg 1$. According to the relation $qc/2\pi a_0= 10^{-2} $Hz, the maximum value of the wave number corresponding to the presence of $m=3$, according to Figure \ref{L2} is equal to $10^{5}$. This value is the upper limit for the detector frequency with the value $10^{-3}$ Hz such that $\kappa^{'}= 5.6 \times 10^{12}/\Omega_{M}h^{2} \gg 1$; which is still a large value\cite{Komatso}. \\
Therefore, this effect might be detected by space-borne methods such as LISA \cite{2017arXiv170200786A}, which would be achieved with a specific detector configuration that targets low-frequency sources such as extreme mass ratio inspirals (EMRIs) \cite{Amaro-Seoane:2010dzj,Aharon:2016kil}. This binary system contains a massive black hole that merges with its smaller companion, such as the white dwarf, neutron star, and stellar mass black hole. The EMRI merges emit GWs at frequencies between $10^{-4}$ Hz to $10^{-1}$ Hz, to which LISA can also be covered. 
 The remarkable point is that this phenomenon can still be investigated with the pulsar timing arrays (PTA) method using NANOGrav projects. Because in wave number values less than $10^{5}$, e.g., $q=10$, the corresponding frequency will be of order $10^{-7}$Hz, which is in the frequency domain of NANOGrav and also keeping that still $\kappa^{'} \gg 1$. Therefore,  the effect of DE on the gravitational waves background can be studied more deeply by observing this stochastic signal ( NANOGrav 15-year data set).
\section{Conclusion}
In this paper, we reviewed all dynamical models of the cosmological constant, $ \Lambda $, as a DE in the equation of gravitational waves.
Although DE is present alongside GWs after the inflation epoch, its energy density is small in the early universe. From this point of view, comparing DE as a function of scale factor, $ a^{-m}, $ with the other models, reveals that its power parameter range (which was determined based on cosmological data) is more limited. The explanation is that for $ 2< m\leqslant 3 $ and even for lower frequencies (or long wavelengths), the upper limit is reduced from $3$ to about $2.8$. In the next epoch, this DE model reduced amplitude, but small wave numbers played an opposite role in contrast to the other models, such as the total vacuum contribution model and explicit function of time. \\
The $\Lambda$-DE model based on time behaves almost similarly to models RVM and GRVS because their reduction amplitudes are $0.84 \dfrac{\sin u}{u^2}$, $0.83 \dfrac{\sin u}{u^2}$ and $0.85 \dfrac{\sin u}{u^2}$, respectively. Here, the amplitudes are very close to each other. 
Comparing all DEmodels revealed that the total vacuum contribution has the maximum reduction of GWs amplitude as $0.63 \dfrac{\sin u}{u^2}$, and the minimum reduction belongs to the GRVS model. Therefore, considering the tensor contribution to the temperature multipole coefficients $C_{l}$ and all of the $ \textquoteleft\textquoteleft$B-B$\textquotedblright$ polarization multipole coefficients, quadratic effects of the tensor modes in the CMB are maximum $60\%$ less than they would be in the case without the damping due to DE terms with total vacuum contribution. The minimum reduction for polarization multipole coefficients is observed in the GRVM, which is $42\%$. 
.\\
Although tensor perturbation strong enough to be detected directly at short wavelengths would also be detected indirectly at longer wavelengths through its effect on the polarization of the cosmic microwave background, the presence of DE can directly affect gravitational waves. Furthermore, it can be seen directly at long wavelengths by Pulsar Timing Array projects, thereby maximizing sensitives to a very low-frequency hum of colliding supermassive black holes. It also directly affects these waves at low frequencies. Accordingly, the sensitivity of the space-borne detector would be comparable to LISA and eLISA.\\
Therefore, the recent announcement of a stochastic signal, the NANOGrav 15-year data set, consistent with a stochastic gravitational waves background, is a great opportunity to prove the imprints of all DE models. Also, the accommodation of nonzero spatial curvature with all DE models has almost no effect on GWs.

\end{document}